\documentstyle[amsfonts,aps,prb,twocolumn]{revtex}
\begin{document}
\draft
\author{Horacio M. Pastawski and Jos\'{e} A. Gasc\'{o}n}
\address{Faculdad de Matem\'{a}tica, Astronom\'{\i}a y F\'{\i}sica, Universidad\\
Nacional de C\'{o}rdoba, \\
Ciudad Universitaria, 5000 C\'{o}rdoba, Argentina}
\title{NMR lineshape in Metallic Nanoparticles: a Matrix Continued Fractions
evaluation}
\date{2 december 1996 }
\maketitle

\begin{abstract}
In metallic nanoparticles, the different electronic environment seen by each
magnetic nucleus produces a distribution of Knight shifts of the NMR
frequencies which is observed as an inhomogeneously broadened lineshape. We
study the fluctuations in the local density of states for $s$ electrons at
the Fermi energy in a simple LCAO model. We resort to a Matrix Continued
Fractions calculation of the Green's functions. Results show that line
broadens asymmetrically and its shift decreases as the particle size or the
temperature diminish satisfying a universal scaling function. However, for
very small particles, surface states become relevant to determine a
lineshape that departs from the universal scaling behavior. These trends are
consistent with the observed tendencies in Cu and Pt particles.
\end{abstract}


\section{Introduction}

Small clusters of atoms of metallic elements have interesting physical
properties\cite{Halperin} which originate in the strong interference effects
that appear when the coherence length $L_\phi $ of the electronic
excitations is larger than the size $L$ of the particle. As already noted by
Kubo\cite{Kubo}, this would allow the manifestation of the geometry
dependent quantization of electron energies. This condition leads to the
very active field of mesoscopic physics\cite{mesoscopic} which involves a
deep revision of traditional ways of calculation and interpretation of
transport and spectroscopic\cite{MBeats} properties. On the other hand the
diverse technical applications of metal particles as well as composites of
metal particles has pushed the development of a wealth of sample preparation
techniques\cite{Halperin}\cite{Brom} which allow varied system
characteristic. It has been recognized\cite{Knight}\cite{MakowkaSlichter}%
\cite{Brom} that NMR can be a useful tool to characterize metallic
nanoparticles and that further insight in the comprehension of their
properties requires the development of models which allow a quantification
of spectra. Such a model should contain the basic physics involved: Each
magnetic nucleus is sensitive to {\it local} magnetic polarization of
electrons which changes the effective field. This manifests\cite{Slichter}
as different (Knight) shifts in the NMR absorption frequencies leading to
inhomogeneous line broadening and different relaxation times. While it is
obvious that a close connection exists between the output of an NMR
experiment and the geometry and disorder of the particle on which the
electronic properties depend, we are still far from a universal quantitative
description of the effect.

In this paper we address the problem of how the size of a particle
contributes to the NMR lineshape. Because of quantum interferences, the
details of shape and disorder of the nanoparticles manifest\cite{Zharekeshev}%
\cite{Batsch} as a Wigner-Dyson or Poisson distribution of electronic energy
levels depending on whether the states are extended (metallic phase) or
localized (dielectric phase). In both cases, there are strong fluctuations
in the local density of states (LDOS) and hence in the local Pauli
susceptibility. Since samples used in earlier experiments had a distribution
of particle sizes, the inhomogeneous line shapes were assigned to this
distribution. Different sizes would not only give a different number of
electrons at the Fermi energy (average density of states) but also involve a
fluctuation in the parity of total electron number giving a further
contribution to the inhomogeneity. However, these efforts, based in a
canonical ensemble, do not give a consistent explanation for the observed
lineshape valid for all the systems studied. A recent {\it break-through}
was proposed\cite{Efetov} with the use of grand canonical ensemble which
allows for fluctuations in the particle charge. This picture, which is valid
in many experimental situations, emphasizes the fact that inhomogeneities
manifest already in a single particle. Therefore inhomogeneous broadening
occurs even when particles are exactly equal. This is observed \cite{Brom}
for the ligand-stabilized metal-cluster compounds Pt$_{309}$Phen$_{36}^{*}$O$%
_{30}.$ Estimations of this shape were performed using an electronic
structure represented by a non-linear sigma model\cite{Efetov} and by a
Random Matrix Theory (RMT)\cite{Beenakker}. Both models produce the same
results for the LDOS fluctuations. One could visualize the model Hamiltonian
implicit in the RMT theory, considering a set on $M$ ''atomic orbitals''
whose interactions, not constrained to a given lattice, are described by the
Hamiltonian matrix elements ${\cal H}_{i,j}$, which are random and satisfy a
gaussian distribution $P({\cal H})=C\exp [-{\rm Tr}{\cal H}_{}^2/(2\sigma
^2)],$ with $C$ a normalization constant and $\sigma ^2\sim M^{}$ to keep a
finite band width $B$. Since the analytical solutions are obtained in the
limit $M\rightarrow \infty $, each ''orbital'' sees an equivalent ''mean
field'' environment. While solutions describe fluctuations in the LDOS, they
can not distinguish surface from bulk ''orbitals'' or the explicit geometry
of the particles neither describe the transition to the localized regime.
Hence, while these analytical models contain the essential physics of level
statistics, they are not expected to provide a detailed characterization of
different materials and particle geometries. In order to overcome this
limitation, in this work we use a very simple tight binding model for the
electronic structure that is a first step toward a more precise description
of the nanoparticles in terms of Linear Combination of Atomic Orbitals
(LCAO). A similar model was used recently\cite{Zharekeshev}\cite{Batsch} to
describe the cross over between Wigner-Dyson and Poisson statistics. We will
also show that resorting to appropriate calculation tools (e.g. Matrix
Continued Fractions) such models can be used in practical calculations.

\section{The Model}

The electronic system is described by the Hamiltonian

\begin{equation}
{\cal H=}\sum_{i=1}^ME_ic_i^{+}c_i^{}+\sum_{j>i}^M%
\sum_{i=1}^M(V_{j,i}c_j^{+}c_i^{}+V_{i,j}c_i^{+}c_j^{})  \label{H}
\end{equation}
where $E_i$ is the energy of the $s$-like atomic state centered at nuclei
position ${\bf r}_i$ orthogonalized atomic orbital $\varphi _i({\bf r}%
)\equiv \varphi ({\bf r-r}_i)$, and $V_{j,i}$ are real numbers accounting
for the kinetic energy of the electrons. There are $M=\Omega /a^3$ orbitals
in the volume $\Omega =L_1L_2L_3.$ To model the disorder we assume that $E_i$%
's are uniformly distributed around the mean value $E_S$ in the range $%
[-W/2,W/2]$ (Anderson model). In a Hartree mean field description, charge
transfer energies are accounted for conserving the structure of (\ref{H}).
The presence of the magnetic field $H$ leads to the unitary symmetry and can
be included in the Hamiltonian through the Peierls substitution. However,
for experimentally accessible fields ($H\leq 10^5{\rm G}$) and typical
particle sizes ($L\geq 10^{-7}{\rm cm}$) its effect is negligible ($%
HL^2e/\left( ch\right) $ $\ll 1$ with $e/\left( ch\right) =$ $2.4\times 10^6%
{\rm G}^{-1}{\rm cm}^{-2}$) because degeneracies are already broken by the
disorder and we restrict our calculation to orthogonal symmetry. In
principle, interactions with other particles produce a decay of the
eigenstates and enter through an inhomogeneous broadening of the ionization
energy of orbitals at surfaces\cite{GLBE} ($E_i\rightarrow E_i-{\rm i}\Gamma
_i$), accounting for electron tunneling toward neighbor particles. While
this is not much complication in a numerical treatment of the problem, to
allow for comparison with previous works, we adopt here a simpler
description in which $\Gamma _i\equiv \eta _o$ at every site. Properties of
the single particle excitation spectrum are contained in the retarded
(advanced) Green's function 
\begin{equation}
G_{i,j}^{R(A)}(\varepsilon _{}^{})=\sum_k\frac{a_k(i)a_k^{*}(j)}{\varepsilon
+[(-){\rm i}\eta -E_k]},  \label{GR}
\end{equation}
where $\psi _k({\bf r})=\sum_ia_k(i)\varphi _i({\bf r})$ and $E_k$ are the
exact eigenfunctions (molecular orbitals) and eigenenergies for the isolated
particle. $\eta $ is a natural broadening of the electronic states and the $%
(-)$ sign corresponds to the retarded Green's function. The local densities
of states per atom (LDOS) at $i$-th site is evaluated as:

\begin{equation}
N_i(\varepsilon )=-(2\pi {\rm i})^{-1}[G_{i,i}^R(\varepsilon
)-G_{i,i}^A(\varepsilon )],  \label{Ni}
\end{equation}
from which the relevant contribution to the density of states per unit
volume at the $i$-th nucleus, $N(\varepsilon ,{\bf r}_i),\,$is obtained. In
presence of an external magnetic field, electrons polarize proportionally to
a {\it local} Pauli susceptibility 
\begin{equation}
\chi _P({\bf r}_i)=2(g\mu )^2\int N(\varepsilon ,{\bf r}_i)[-\frac{\partial 
{\rm f}}{\partial \varepsilon }]{\rm d}\varepsilon ,  \label{pauli}
\end{equation}
with ${\rm f}(\varepsilon )$ is the Fermi function. Numerical integration of
(\ref{pauli}) for finite temperatures can be very time consuming due to the
singular nature of the spectrum which would require a LDOS calculation for
each energy. In order to avoid integration, the finite temperature effects
can be taken into account by taking 
\begin{equation}
-[\frac{\partial {\rm f}}{\partial \varepsilon }]\approx \frac 1\pi \frac{%
a_Tk_BT}{(\varepsilon -\varepsilon _F)_{}^2+(a_Tk_BT)^2}\,.  \label{lorentz}
\end{equation}
which in practice becomes equivalent to put $k_BT=0$ in Eq.\ref{pauli} and
use an energy uncertainty for the electronic levels

\begin{equation}
\eta =\eta _o+k_BT,  \label{eta}
\end{equation}
where the coefficient $a_T\equiv 1$ gives a good approximation approximation
for Eq. (\ref{lorentz}). The Knight shift at a given nucleus is 
\begin{equation}
\Delta \omega ({\bf r}_i)=J\frac 1{g\mu }\chi _P({\bf r}_i)H,  \label{Ki}
\end{equation}
where $J$ is the hyperfine constant and $g\mu $ the electron magnetic
moment. Then, $\Delta \omega ({\bf r}_i)$ will follow a probability
distribution proportional to that of the electron LDOS. The NMR absorption
line is inhomogeneously broadened by this distribution and is obtained by
summing up all magnetic nuclei. The resonance line finally becomes 
\begin{equation}
I(\omega )=A\sum_i\delta [\omega -\omega _0-\Delta \omega ({\bf r}_i)],
\label{I}
\end{equation}
where $\omega _0$ is the non-shifted position of the resonance in a
dielectric material (as in a salt). That is, the observed NMR line shape is
the sum of narrow lines with and inhomogeneously distribution in frequencies
that follows that of the LDOS. In order to allow comparisons between
systems, it is convenient to define the adimensional frequency variable $%
x=(\omega -\omega _o)/(\omega _K-\omega _o),$ where $\omega _K$ is the peak
resonance in the bulk metal. This gives $x=1$ for a metal and $x=0$ for the
dielectric phase$.$

The intensity can be normalized: $\int I(x){\rm d}x$ $=1$. A quantity of
interest to characterize the line is the Knight shift $x_m$ defined as $%
I(x_m)=\max [I(x)];$ another is the first moment $x_1=\int xI(x)\,{\rm d}x,$
whose difference from $x_m$ is a measure of the asymmetry of the line.
Notice that if we are working within a grand canonical ensemble in an
infinite system, we expect that even when an electronic density at some
nuclei is diminished by interferences this is compensated by an increase in
others. In the analytic models\cite{Efetov} this property results in $x_1=1.$
However in a MO model of the cluster this is not necessarily true. In a
cluster, the ensemble average DOS is not a monotonic function of energy but
an oscillating function with local maxima at the eigenenergies of the
ordered cluster. This dependence on on the cluster size is an easy
analytical result\cite{Rodrigues} for a Lorentzian distribution of site
energies.

\section{Matrix Continued Fraction}

The exact solution of Hamiltonian (\ref{H}) is a formidable task. Rough
features of the LDOS can be obtained by the Recursion Method\cite{HHK}. This
involves the obtention of a tridiagonal basis for (\ref{H}) and the
evaluation of the local Green's function (\ref{GR}) as a continued fraction.
When used within the RMT this method allows analytical solutions\cite{Mattis}%
. We resort to it for the evaluation of the density of states per orbital in
an infinite ordered system, $N_0(\varepsilon _F,\eta )$. From this quantity
the typical level separation for a cluster with $M$ orbitals is calculated:

\begin{equation}
\Delta =1/[M\cdot N_0(\varepsilon _F,\eta )].  \label{Delta}
\end{equation}
However, for small $\eta $ (low temperature)$,$ it is difficult to obtain
the precise fluctuations in the LDOS for finite clusters, because the
intrinsic numerical instabilities\cite{Haydock}\cite{Cullum-Willoughby} of
the tridiagonalization procedure distort the fine details of the LDOS\cite
{AAQA}. It is therefore compelling to resort to a more robust method to
describe the infinitesimal details of the density of states. This is the
Matrix Continued Fractions (MCF) method\cite{MCF} for the calculation of the
local Green's function. Its basic idea is to exploit the short range
interactions in Hamiltonian (\ref{H}) by indexing states in a way that
subspaces representing layers interact through nearest neighbor subspaces.
In matrix form:

\begin{equation}
{\cal H=}\left[ 
\begin{array}{ccccc}
\ddots & \ddots & {\bf 0} & {\bf 0} & {\bf 0} \\ 
\ddots & {\bf E}_{n-1,n-1} & {\bf V}_{n-1,n} & {\bf 0} & {\bf 0} \\ 
{\bf 0} & {\bf V}_{n,n-1} & {\bf E}_{n,n} & {\bf V}_{n,n+1} & {\bf 0} \\ 
{\bf 0} & {\bf 0} & {\bf V}_{n+1,n} & {\bf E}_{n+1,n+1} & \ddots \\ 
{\bf 0} & {\bf 0} & {\bf 0} & \ddots & \ddots
\end{array}
\right] ,  \label{HM}
\end{equation}
where ${\bf 0}$'s are null matrices, ${\bf E}$'s in the diagonal represents
intra-layer interactions while the only non-zero off-diagonal matrices and $%
{\bf V}_{n,n\pm 1}$ connect nearest neighbor layers. Detailed structure of
the submatrices depends on the lattice, for the cubic structure ${\bf V}%
_{n,n\pm 1}=V{\bf 1}$, with ${\bf 1}$ the identity matrix. The local
retarded Green's functions connecting sites $i$ and $j$ within the $n$-th
layers are arranged in a matrix 
\begin{equation}
{\bf G}_{n,n}^R\left( \varepsilon \right) =\left[ \left( \varepsilon +{\rm i}%
\eta \right) {\bf 1}{\Bbb -}{\bf E}_{n,n}-{\bf \Sigma }_n^{R\,+}\left(
\varepsilon \right) -{\bf \Sigma }_n^{R\,-}\left( \varepsilon \right)
\right] ^{-1}
\end{equation}
where the matrix self energies ${\bf \Sigma }_n^{R\,+}$ and ${\bf \Sigma }%
_n^{R\,-}$ are calculated in terms of Matrix Continued Fractions (MCF)
defined through the recurrence relations: 
\begin{equation}
{\bf \Sigma }_n^{R\,\,\pm }={\bf V}_{n,n\pm 1}\frac{{\bf 1}}{\left(
\varepsilon +{\rm i}\eta \right) {\bf 1}{\Bbb -}{\bf E}_{n\pm 1,n\pm 1}-{\bf %
\Sigma }_{n\pm 1}^{R\,\,\pm }}{\bf V}_{n\pm 1,n},
\end{equation}
which are calculated with the boundary conditions: ${\bf \Sigma }%
_{L_3}^{+}\equiv {\bf \Sigma }_1^{-}\equiv {\bf 0}.$ The intrinsic stability
of the method manifests\cite{Slutzky} in the fact that a modification of a
layer at distance $L$ away from a given one will produce an exponentially
small variation of the local density of states, $\delta N\sim \exp \left[
-\gamma _1L\right] .$ Here $\gamma _1$ is the minimum Lyapunov
Characteristic Exponent. In the localized regime ($W>W_C$, $\eta =0$) $%
1/\gamma _1$ coincides with the localization length while in the ordered
case ($W\ll W_C$, $\eta >0$) it gives a phase breaking length\cite{Rodrigues}
$L_\phi (\eta )$ which washes out the finite size effects when $L>L_\phi $.

\section{Results}

In this work we consider a simple cubic structure forming clusters of cubic
shape with sizes $L$ ranging from 5 to 15 layers. That makes clusters of up
to 3375 orbitals. We chose an arbitrary Fermi energy $\varepsilon
_F=E_S+4\left| V\right| $ which is far enough from the van Hove
singularities for that lattice and hence allows a comparison with results
from the non-linear sigma model which implies nearly free electrons.
However, with $-V\approx 1e{\rm V=1.16\times 10}^4{\rm K,}$ it gives a
reasonable scale to describe an $s$ metal such as Cu. Disorder is small
enough ($W=0.5\left| V\right| $) to be on the metallic side of the metal
insulator transition but strong enough to provide randomness of the energy
levels. Further sampling is achieved by considering an ensemble of 10 Fermi
energies in a range $\delta \varepsilon _F=0.5\left| V\right| $. This
minimizes the effect of finite size correlations in the electron energy
spectrum.

Figure 1 shows the normalized distribution of LDOS for a cluster of size
L=15 with three temperatures producing effective broadenings $\eta /V=0.05$, 
$0.0067$ and $0.001$; which according to Eq. (\ref{eta}) can be interpreted
as a sequence of decreasing temperatures. The number of occurrences $I(x)$
of each normalized density of states 
\begin{equation}
x=\frac{N_i(\varepsilon _F,\eta )}{N_0(\varepsilon _F,\eta )}  \label{x}
\end{equation}
is proportional to the NMR absorption. Each member of the ensemble,
containing four disorder configurations, is normalized according to the
corresponding Fermi energy and line broadening, a dependence writen
explicitly in Eq.(\ref{x}). The same results were obtained with a single
cluster, since the system is big enough to provide a configurational
average. It is clear that by lowering the temperature the Knight shift
decreases from its metallic limit $x_m=1$ toward $x_m=0$ corresponding to a
salt. Simultaneously the line becomes increasingly asymmetric. These results
are consistent with the experiments in Cu \cite{Knight}\cite{Kobayashi} and
Pt\cite{MakowkaSlichter}.

In Fig. 2 we show the variation of the lineshape with cluster sizes (L=10, 8
and 6) for a fixed natural broadening of $\eta /V=0.018$. Again it confirms
the general trend experimentally observed in Cu and the predictions of the
analytical models, i.e. the band tends to broadens and maxima shift to lower
frequencies as particle size or temperature is decreased.

In order to capture the universal nature of the competition between cluster
size and temperature suggested by Figs. (1) and (2), we plot in Fig. (3) the
value of the adimensional Knight Shift as a function of the scaling
parameter $\eta /\Delta \equiv \alpha /(2\pi )$ which measures the ratio
between the level broadening and the typical level separation. $\alpha $ is
the scaling parameter defined in ref. [\cite{Efetov}]. For completeness we
also include in the plot a continuous curve obtained from the analytical
lineshape obtained in that work for the unitary ensemble. As expected for a
given $\eta /\Delta $ their results tend to give a higher $x_m$ than our
numerical values. This is consistent with the fact that the breaking of time
reversal symmetry produces a stronger level repulsion diminishing the
fluctuations.

However, there is no universal scaling function that fits the complete
lineshape for all systems. This is because in our model there are surfaces.
Although we did not assign different ionization energies to surface orbitals
(which can produce Tamm states), local densities of states at surface region
(about $6/L$ of the total) can differ substantially from that at inner atoms%
\cite{Schrieffer}. Sites at the surfaces are connected to fewer neighbors
and this manifest on a decrease in the wings of the LDOS as a consequence of
lower gain in kinetic energy\cite{Dpolish}, which is compensated with an
increase of the LDOS at energies around $E_S$. It should be emphasized,
however, that the sign of the change in the LDOS\ at surfaces depends both,
on the lattice structure and on the value of $\varepsilon _F$; e.g. if $%
\varepsilon _F=E_S$ the surface LDOS would be higher than its value in the
bulk. In any case, this produces an additional fluctuation in the LDOS which
is more important for smaller systems. These effects attenuate within a few
lattice constants and are consistent with the exponential healing length
model \cite{vdKlink}. In Fig. $(4)$ we separate the contributions to the
distribution of LDOS (linewidth) from surface and inner sites in a small
cluster. Similar separations of maxima are obtained at higher $\eta /\Delta
. $ To emphasize the importance of this effect on the line broadening in
small particles, we show in Fig.(5) the scaling function satisfied by $x_m$
as function of the adimensional parameter $\eta /\Delta $ for a $5\times
5\times 5$ cluster in which we also discriminate the contributions of
surface and inner sites$.$ It is clear a general tendency toward lower
values of the Knight shift for nuclei at the surfaces. Maxima do not
coincide even when $\eta /\Delta $ is increased. In small clusters this
effect manifests as an intrinsic linewidth persisting at high temperatures,
a tendency that departs from the scaling law. This is the situation observed
in recent experiments\cite{vdPutten}\cite{Brom96} on Pt$_{309}$ and Ni$_{38}$%
Pt$_6$ particles, where linewidths do not shrink with increasing
temperature. Our results suggest that size and temperature dependence of
line shapes can be used in conjunction with other NMR techniques such as
SEDOR\cite{MakowkaSlichter} to allow the experimental identification of
electronic densities at surface regions.

\section{Final Remarks}

, In agreement with analytical theories\cite{Efetov}\cite{Beenakker} for
mesoscopic fluctuation, our results show that distribution of LDOS (and
hence NMR line) shifts and asymmetrically broadens as temperature decreases
and particle size diminishes and can be described by a universal scaling law
which is consistent general trends observed in Cu particles and the size
dependence \cite{vdPutten}\cite{Brom96} of Pt$_{309}$ and Ni$_{38}$Pt$_6$
particles, although the origin of the disorder in the last cases is not
quite clear. In this case a finite broadening is obtained even for high
temperatures. These departures from scaling theory in very small particles,
is understood in our model as due to the presence of surfaces, a feature
that the analytical theories do not contain. Therefore, our LCAO model
contains the main features of mesoscopic fluctuations and models the
exponential healing of surface effects. Due to the simplicity and speed of
the MCF calculation, it allows for further improvements of the model
Hamiltonian which should lead to a more detailed evaluation of the
electronic structure. Also orbital susceptibility could be quantified within
these models. Since Cu has also a contribution to the linewidth due to
quadrupolar interactions, and Pt, having a $d$-band contribution to the
Fermi surface, has an important core polarization susceptibility, they are
far from ideal systems. A good candidate for a full quantification the
described behavior is Ag, with spin 1/2 nucleus and $s$ electrons.
Therefore, further development along these lines would allow for: 1) a
better quantification of NMR experiments; 2) the investigation of the
electron states determining the chemical kinetics at surfaces and 3) the
progress in the understanding of basic phenomena at a mesoscopic scale.
While the first issue was the main one addressed in this article we would
like to add a few comments on the other two. Since in a frontier-orbital
picture of bonding to metal surfaces, the states around the Fermi energy are
those that determine the chemical reactions, it is of great interest to
determine how these are modified by changing the size and disorder of the
catalyst particle\cite{Hoffmann}. In particular a metal-insulator transition
in the low dimensional (Tamm) surface band could play a very relevant role
in chemical kinetics. Therefore, the theoretical and experimental methods
described in this paper can be conveniently adapted to suit this
investigation. {\bf \ }Finally, the interference phenomena lying in the core
of mesoscopic physics could be controlled to obtain a fine tuning of many
static and dynamical properties of materials. In particular, the present
technique could be used to complement current studies\cite{Batsch} that
attempt to quantify the interplay of disorder, magnetic field and
temperature in the metal-insulator transition.

\section{Acknowledgments}

This work was performed at LANAIS de RMN (UNC-CONICET) with financial
support from Fundaci\'{o}n Antorchas, CONICOR and SeCyT-UNC. HMP is a member
of CONICET and appreciates discussions with H. Brom, C. Beenakker, V.
Prigodin and K. Efetov.

\section{Figure Captions}

\bigskip\ \noindent {\bf Figure 1}. Typical distribution of occurrences $I$
of LDOS (or NMR lineshape) as function of the adimensional frequency $%
x=(\omega -\omega _0)/(\omega _K-\omega _0)$ $=N(\varepsilon _F,\eta
)/N_0(\varepsilon _F,\eta )$ evaluated for an ensemble of 4 clusters of $%
15\times 15\times 15$ orbitals sampling 10 energies around $\varepsilon
_F=E_s+4\left| V\right| $ at different temperatures: $\eta /V$ $=\eta
_o+k_BT $ $=0.05,$ $0.0067$ and $0.001$ The disorder is $W=0.5\left|
V\right| $.

\bigskip\ \noindent {\bf Figure 2}. Variation of NMR lineshape $I$ for an
ensemble of 10 particles of sizes $L=6,$ $8$ and $10$ lattice units$,$ with
10 energies each. Temperature is such that $\eta =0.018\left| V\right| ,$
and other parameters are as in Fig. (1).

\bigskip\ \noindent {\bf Figure 3}. Scaling curve for the Knight shift $x_m$
for particles of different sizes as function of the adimensional parameter $%
\eta /\Delta .$ The continuous curve is obtained from the analytical
solution for a unitary ensemble of ref.\cite{Efetov}.

\bigskip\ \noindent {\bf Figure 4}. Surface and bulk contribution to the
line shape in a small cluster ($M=5\times 5\times 5$), calculated with an
ensemble of 20 different disorder configuration and 10 energies each.
Temperature is such that $\eta /V=0.031.$ The dotted plot corresponds to
normalized contributions of orbitals in the surface of the cluster. Full
line plot is from inner sites.

\bigskip\ \noindent {\bf Figure 5}. Surface and Bulk contribution to the
Knight shift (or maximum in the line) is shown here for a small cluster with
varying $\eta /V$ $=\eta _o+k_BT$ evaluated from curves as those in Fig.
(4). Open triangles represent inner nuclei while open circles are those on
the surface. Filled squares represent the Knight shift evaluated from all
sites.

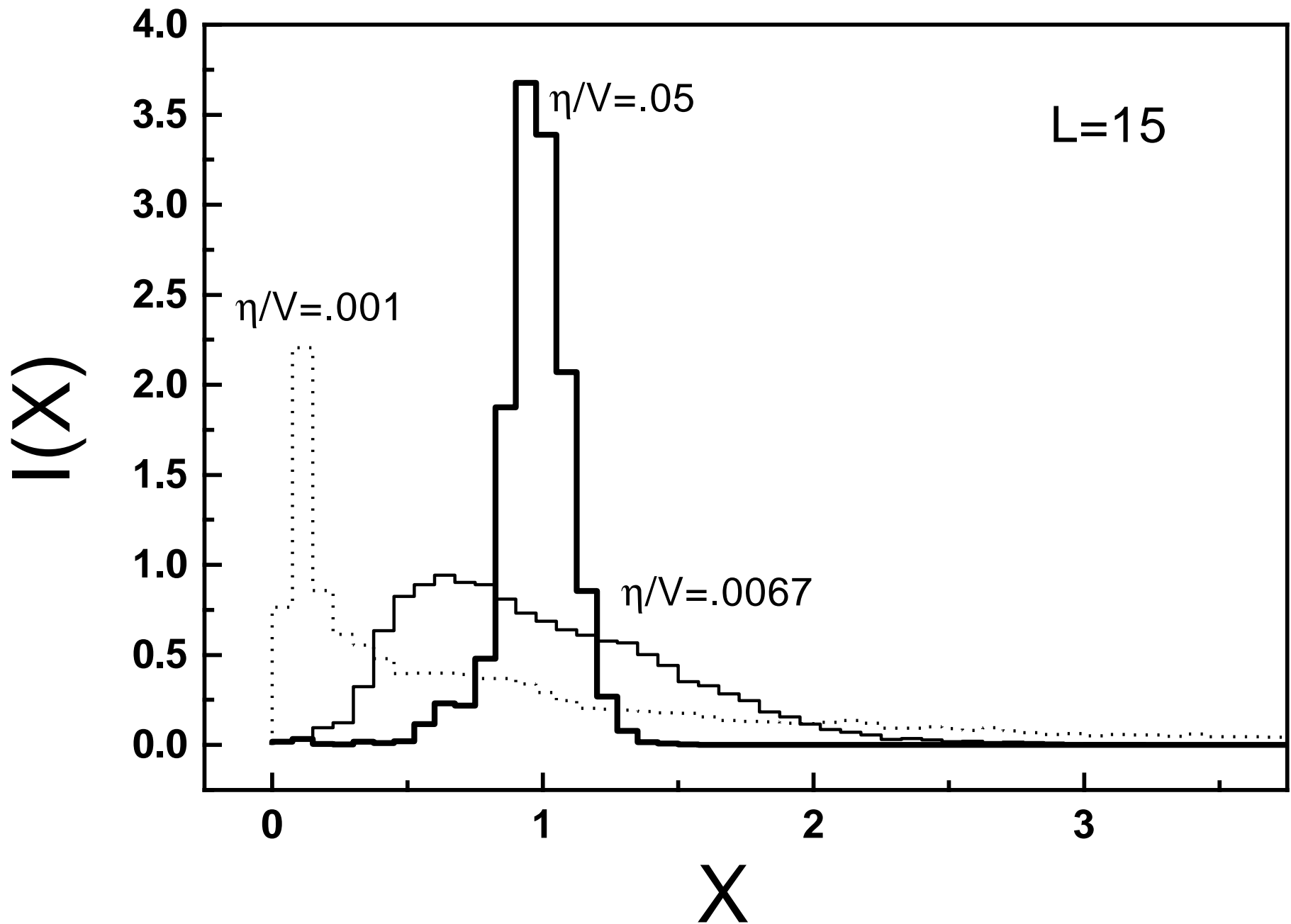
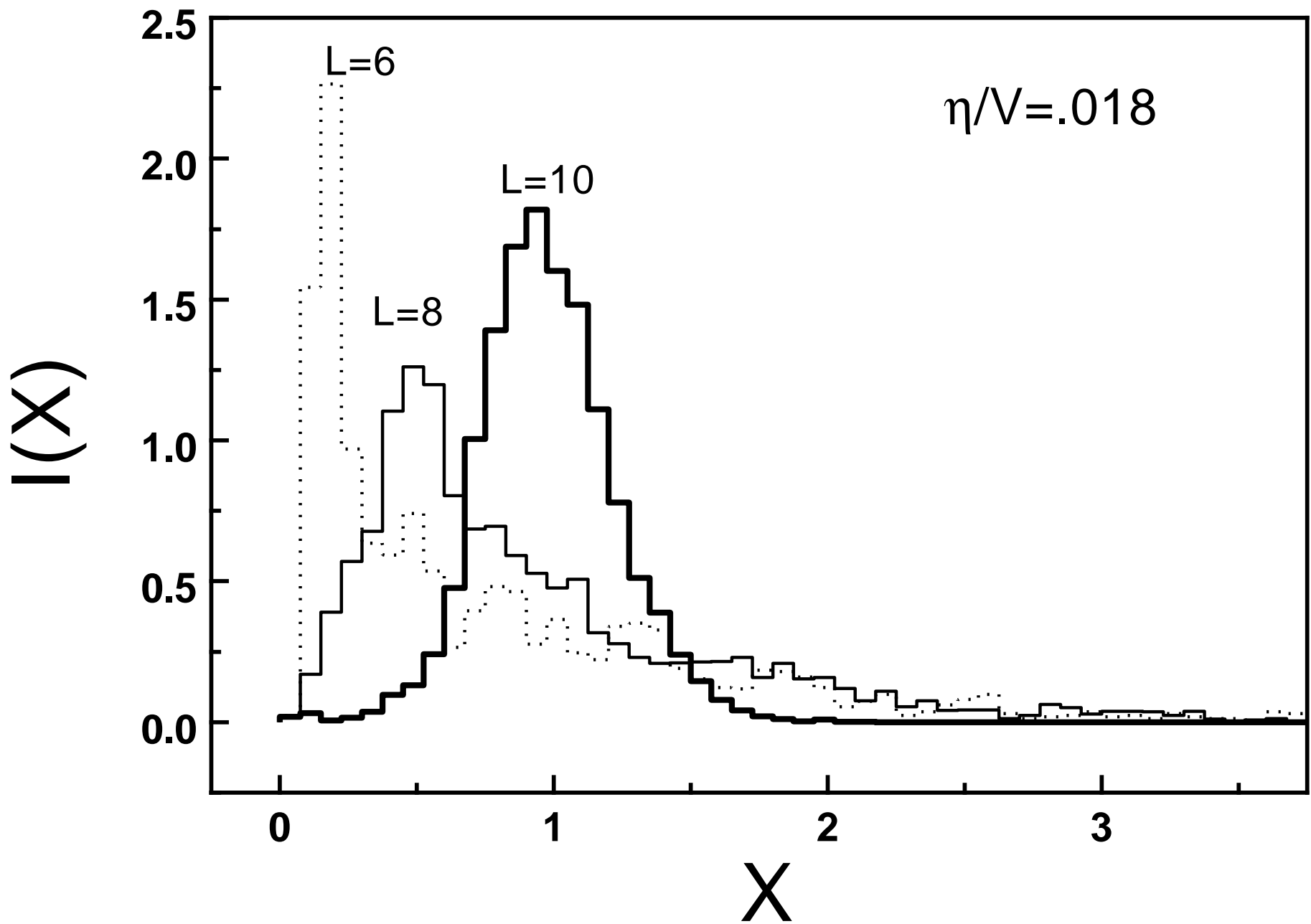
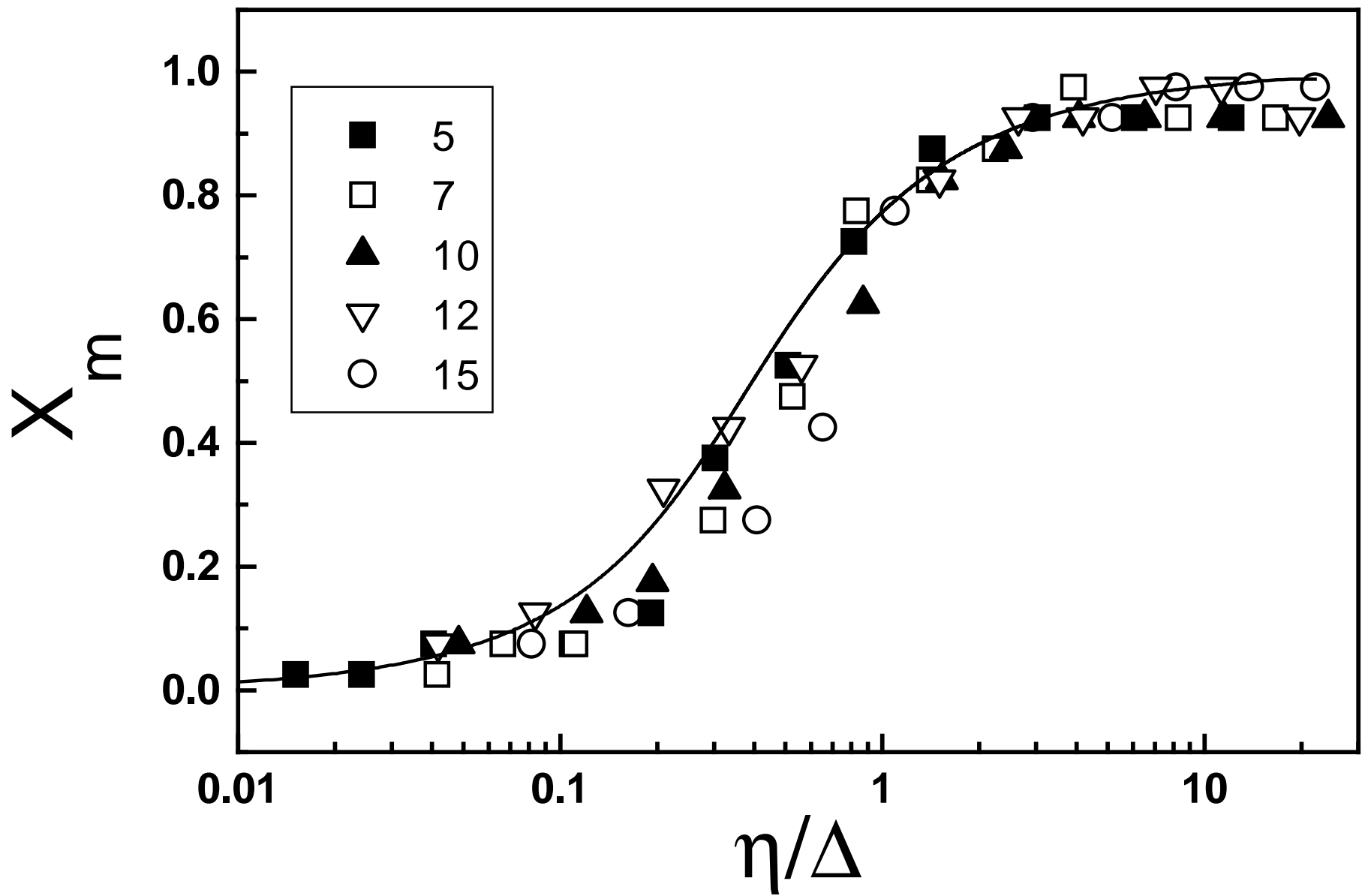
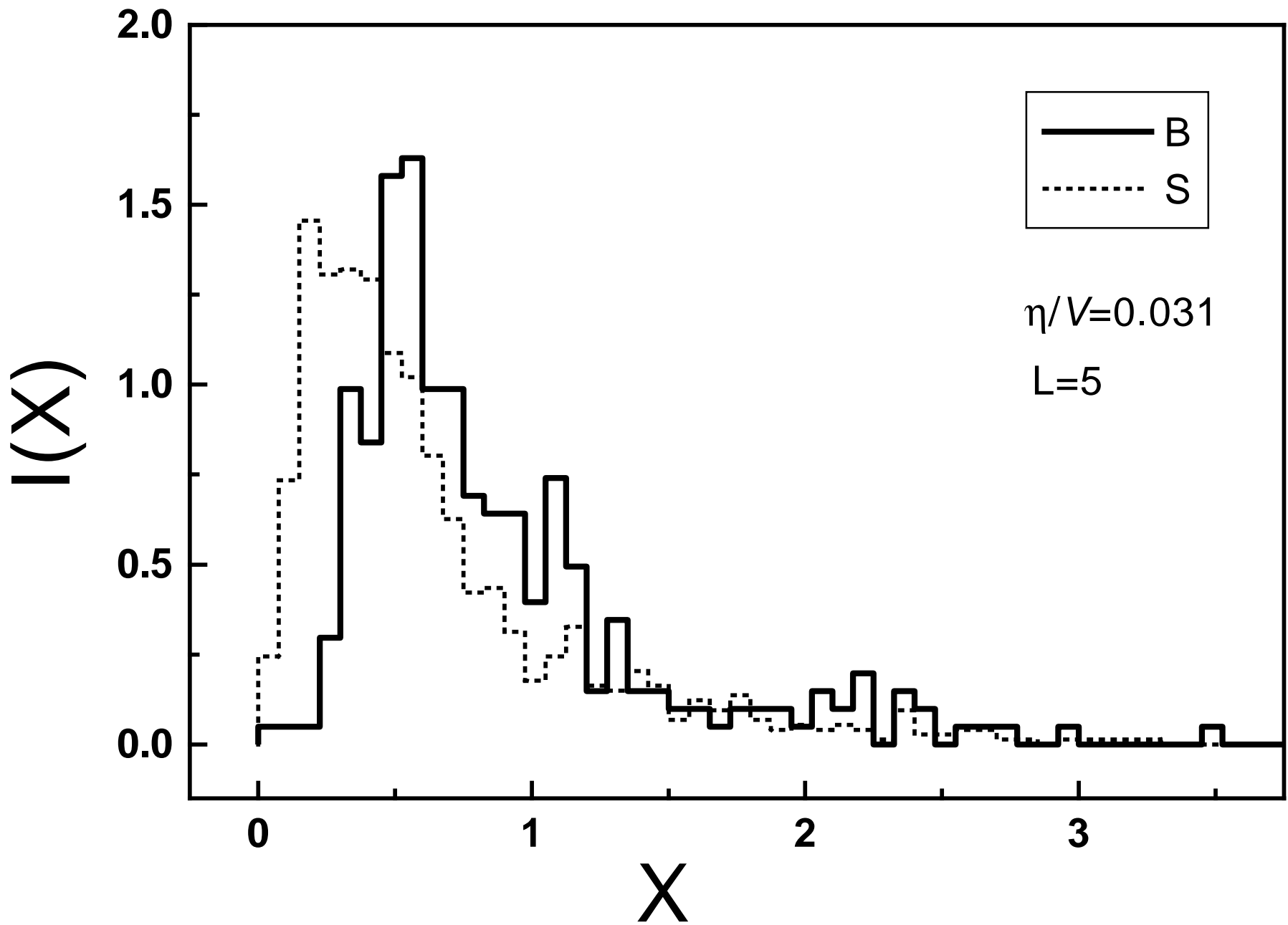
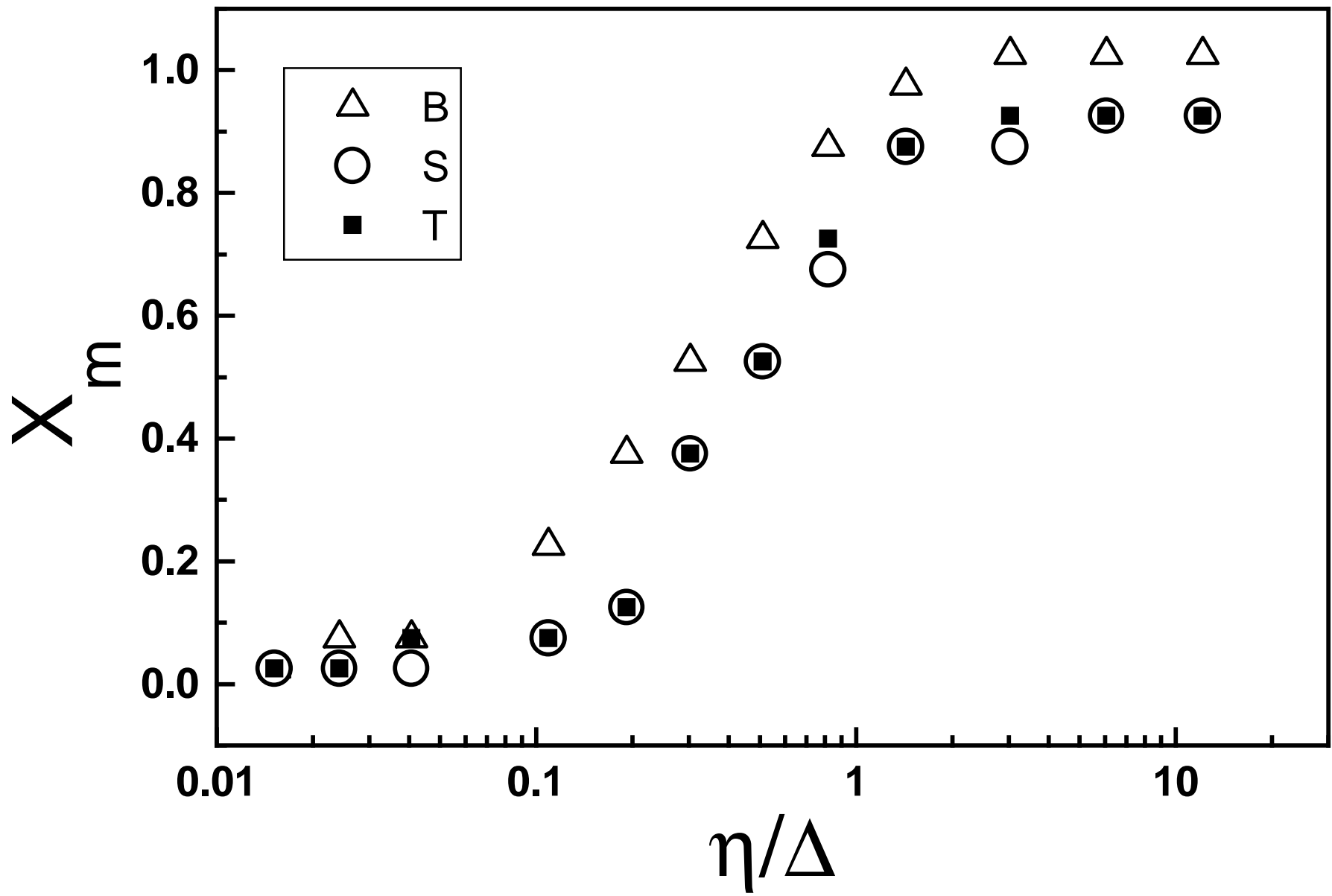
\end{document}